\documentclass[%
 reprint,
superscriptaddress,
 amsmath,amssymb,
 aps,
prb,
]{revtex4-2}

\usepackage{graphicx}
\usepackage{dcolumn}
\usepackage{bm}

\begin{document}

\preprint{APS/123-QED}

\title{Extended Hubbard Model realized in 2D clusters of molecular anions}

\author{Oliver Tong}
\affiliation{Department of Physics \& Astronomy, University of British Columbia, Vancouver Canada, V6T 1Z1}
\affiliation{Blusson Quantum Matter Institute, University of British Columbia, Vancouver Canada, V6T 1Z1}
\author{Katherine A. Cochrane}
\affiliation{Department of Chemistry, University of British Columbia, Vancouver Canada, V6T 1Z1}
\affiliation{Blusson Quantum Matter Institute, University of British Columbia, Vancouver Canada, V6T 1Z1}
\author{Bingkai Yuan}
\affiliation{Blusson Quantum Matter Institute, University of British Columbia, Vancouver Canada, V6T 1Z1}
\author{Tanya Roussy}
\affiliation{Department of Physics \& Astronomy, University of British Columbia, Vancouver Canada, V6T 1Z1}
\affiliation{Blusson Quantum Matter Institute, University of British Columbia, Vancouver Canada, V6T 1Z1}
\author{Mona Berciu}
\affiliation{Department of Physics \& Astronomy, University of British Columbia, Vancouver Canada, V6T 1Z1}
\affiliation{Blusson Quantum Matter Institute, University of British Columbia, Vancouver Canada, V6T 1Z1}
\author{Sarah A. Burke}%
 \email{saburke@phas.ubc.ca}
\affiliation{Department of Physics \& Astronomy, University of British Columbia, Vancouver Canada, V6T 1Z1}
\affiliation{Department of Chemistry, University of British Columbia, Vancouver Canada, V6T 1Z1}
\affiliation{Blusson Quantum Matter Institute, University of British Columbia, Vancouver Canada, V6T 1Z1}

\date{\today}

\begin{abstract}
The Hubbard model, despite its simplicity, is remarkably successful at describing numerous many-body phenomena. However, due to the small class of problems which can be solved exactly, there has been substantial interest in quantum simulations of extended Hubbard models to in turn, simulate materials and the interaction-driven phases they host. Here, we study small clusters of molecular anions of 3,4,9,10-perylene tetracarboxylic dianhydride on NaCl bilayers on Ag(111) using non-contact Atomic Force Microscopy, Electrostatic Force Spectroscopy, and Scanning Tunnelling Microscopy and Spectroscopy, and show that the occupation and transition energies are well described by an extended Hubbard model. In particular, asymmetric clusters of four molecules require the addition of differing inter-site electrostatic interaction terms and on-site potentials, as well as asymmetric hoping terms. With $t<<U$, occupation asymmetry is driven by these terms, independent of U. The good agreement between the model and the data indicate such molecular anion clusters could be used to probe larger systems and a more varied phase space of realistic fermionic Hubbard models.
\end{abstract}


\maketitle

The electronic spectra of the vast majority of materials can remarkably be described considering only one electron at a time. This works when the electrons are highly delocalized and screen one another such that they do not feel the electrostatic interactions with the other 10$^{23}$ or so electrons. However, this approximation does not generically hold true, and indeed we miss some of the most exciting phenomena in the electronic behaviour of materials under such assumptions, including superconductivity and interaction-driven phases such as Mott insulators and many magnetic phases. The Hubbard model\cite{Hubbard.1963} adds a characteristic ``on-site" repulsion of electrons -- i.e. that electrons do not like to occupy the same small space of say, an atomic orbital -- through the addition of the so-called ``Hubbard U" to attempt to capture the influence of the interactions between electrons. This interaction strives to keep electrons apart and tends to drive localization, competing with the kinetic energy  (controlled by the hopping parameter $t$) that characterizes delocalization of electrons in bonded solids. Despite its relative simplicity, the Hubbard model has been remarkably successful in capturing many rich interaction-driven phases.

However, Hubbard models are not generically exactly solvable. Among the simplified systems used to study the behaviour of these models, small clusters have been used as tractable systems to probe the effect of different geometries and model the distortions that are often in part driven by interaction effects. Such small cluster models with 4 or 5 nodes were found to be solvable yielding ground states with different distributions of electron occupation, spin, and tendency towards distortion that depended on the ratio $t/U$\cite{Ishii.1984,Callaway.1986}.

Beyond the realm of the real materials that such models are intended to simulate, there has been recent interest and emerging capability in quantum simulations that are in essence, simplified experimental realizations of various forms of the Hubbard model. Experimental platforms for simulating Hubbard models include optical lattices for both bosonic and fermionic Hubbard models\cite{Dutta.2015}, moir\'{e} superlattices\cite{Tang.2020,Kennes.2021}, quantum dots\cite{Manousakis.2002,Barthelemy.2013,Hsiao.2024}, atomic dopants\cite{Salfi.2016,Dehollain.2020,Kiczynski.2022}, and superconducting circuits\cite{Barends.2015}. These quantum simulators afford a degree of control over the parameters of the Hubbard model and can be used to realize a variety of configurations, including small clusters, revealing previously unobserved correlated states\cite{Dehollain.2020}. However, in working with real systems these models often need to be modified: in optical lattices the interactions are typically dipolar not Coulomb; the influence of additional bands may be non-negligible\cite{Dutta.2015}; and in all cases, inter-site interactions and variations in site potentials are often required to describe reality. These ``extended Hubbard models'' are nevertheless useful and perhaps point to a need to include these contributions in real materials to capture the full richness of behaviour observed in interacting systems. 

While the Hubbard model encodes an interaction that is explicitly local -- demanding local measurement of these interactions, occupation, and spin state\cite{Salfi.2016} -- most measurements of Hubbard physics in materials are made through  measurements of the spectral function of the ensemble. For example, electron spectroscopies like photoemission/inverse photoemission spectroscopy display features in the electron addition and removal spectra due to the Hubbard U\cite{Damascelli.2003}, where these interaction-driven changes to the ensemble spectra are a proxy to the interaction itself which is local, and electrostatic in nature.  

Electrostatic force spectroscopy (EFS) can be used to visualize the local charge distribution and detect charge switching on the single-electron scale\cite{Schönenberger.1990,Woodside.2002,Cockins.2010,Gross.2009,Steurer.2015,Scheuerer.2020}. Pixel-by-pixel EFS is a method of spatially probing and manipulating these local surface charges with submolecular resolution\cite{Mohn.2012,Patera.2025}. If an atom, defect, or molecule undergoes a change in charge during the applied bias sweep, a dip or a jump in the EFS spectrum is observed\cite{Steurer.20145z,Ondráček.2016pq4,Kocic.2015qj,Kocic.2017fhk,Huff.2019}. On bulk insulators\cite{Steurer.2015} or thick NaCl films ($>$5ML)\cite{Steurer.20145z}, these charges are stabilized and can be controlled, while on thin films these changes in charge occupation may only be transient and short-lived, but still detectable. This provides a unique opportunity to probe both the local charge occupation as well as the local electrostatic landscape and structure of a nanoscale assembly exhibiting Hubbard physics on the molecular and even sub-molecular scale. 

Here, we probed small clusters of an organic molecule, 3,4,9,10-perylene tetracarboxylic dianhydride (PTCDA), by scanning tunnelling microscopy (STM), scanning tunnelling spectroscopy (STS) and electrostatic force spectroscopy (EFS). PTCDA has a large electron affinity and becomes a molecular anion when placed on bilayer NaCl on Ag(111)\cite{Cochrane.201593h,Dolezal.2022} making it a good prototype system with singly occupied molecular orbitals to experimentally probe a simple Hubbard-like system. Two types of 4-site clusters were studied in depth with differing geometry and modelled using an extended Hubbard model that incorporates intersite electrostatic potentials and varying site energies in addition to the Hubbard U. Here, $U
\gg t$ and electrons tend towards localization as double-occupation is avoided, yet, a rich landscape of ground and excited states was found both experimentally and theoretically which are controlled not by $U$ but by the inter-molecular electron interactions. As such, this work provides an example where the non-local parts of the extended Hubbard interaction actually control the behavior of the system, even though they are significantly smaller than the local $U$ term.

\subsection{Experimental details}

\begin{figure}
    \centering
    \includegraphics[width=1\linewidth]{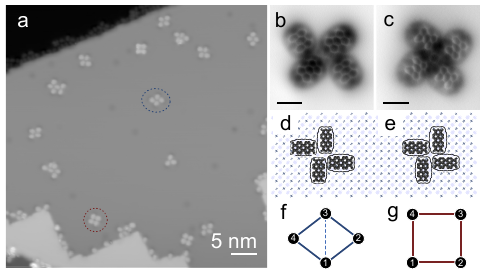}
    \caption{PTCDA clusters on NaCl bilayer on Ag(111). (a) STM topography (50 nm x 50 nm, $I_t$ = 15 pA, $V_b$ = 0.5 V) overview of variety of clusters with several 4-molecule clusters displaying two distinct geometries (circled in red and blue dashed lines). Constant height high resolution ncAFM images taken with CO-functionalized tips of the ``Diamond'' $C_2$ (4 nm x 4 nm, $V_b$ = 0 V) (b) and ``Clover'' $C_4$ (4 nm x 4 nm, $V_b$ = 0 V) (c) clusters exhibiting different symmetries, their structural models in (d) and (e), and connectivity in terms of the Hubbard model in (f) and (g), respectively. }
    \label{fig:overview}
\end{figure}

PTCDA molecules deposited onto clean and freshly prepared NaCl bilayers on Ag(111) at low temperature and subsequently annealed, resulted in small clusters as shown in figure \ref{fig:overview}. The thin insulating NaCl film prevents hybridization between the molecule and metal \cite{Repp.2006h6n}, but allows for tunnelling between the cluster and a reservoir, as well as reducing the scale of U via screening by the underlying metal. Clusters of four molecules were relatively common and displayed two distinct structures: a lower symmetry $C_2$ ``diamond'' shape (see Fig. \ref{fig:overview}b,d), and a $C_4$ symmetric ``clover'' shape (see Fig. \ref{fig:overview}c,e). These two structures are stabilized by quadrupolar intermolecular interactions that favour the anhydride oxygen atoms facing the hydrogen on the aromatic perylene core \cite{Umbach.1998,Würthner.2004}, as well as a strong interaction of the negatively charged anhydride ends of the molecule with the positive Na ions of the surface \cite{Burke.2008ja,Jia.2016}. The two distinct structures feature different local polarization environments \cite{Cochrane.201593h} and different connectivity (see Fig \ref{fig:overview}f,g) and therefore different intermolecular hopping parameters, making them an ideal case study for a small cluster Hubbard model.

Simultaneous STS and electrostatic force spectroscopy measurements were acquired on each type of 4-molecule PTCDA cluster. The STS resembled previous results on PTCDA clusters \cite{Cochrane.201593h}, with a HOMO onset $\sim -1$ V, and shoulder corresponding to the lower Hubbard state, or SOMO, and a upper Hubbard state or SUMO at $\sim +0.5$ V. The nearly degenerate LUMO/LUMO+1 have an onset $\sim 2$V \cite{Cochrane.201593h}. Additional dips in the STS can be seen where jumps occur in the frequency shift data that are not seen in STS acquired with a rigid STM tip, likely due to coupling between the cantilever motion and tunnelling current at the jumps\cite{Klein.2004,Cockins.2010,Kocic.2015qj}. Frequency shift vs. bias curves follow the expected parabolic behaviour but with jumps, or discontinuities, on molecular sites at different biases depending on the site type. These jumps demark segments of the curve, each of which follow parabolic behaviour, but with different offsets indicative of different local surface potential under different regimes of applied bias. For ``clover'' clusters, all sites showed jumps at the same energies, while for ``diamond'' clusters, the two distinct types of sites labelled A (green curves), corresponding to sites 2 and 4, and B (blue curves) corresponding to sites 1 and 3, showed more prominent jumps either on the positive sample bias (electron addition) or negative bias (electron removal) part of the spectrum.

\begin{figure}
    \centering
    \includegraphics[width=1\linewidth]{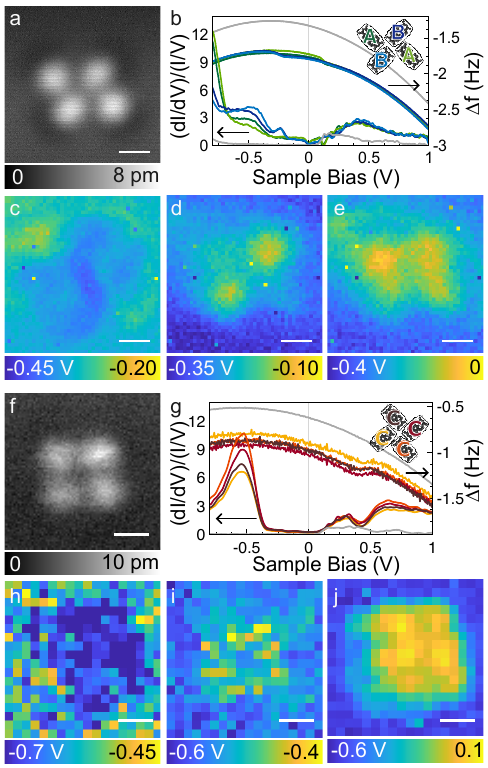}
    \caption{Surface potential measurements of ``Diamond'' (a-e) and ``Clover'' (f-j) islands. (a) topography (b) simultaneous normalized STS (left axis) and $\Delta$f(V) (right axis) spectra of a ``diamond" shaped 4-molecule island taken from constant-height pixel-by-pixel grid with set-point on NaCl(2ML)/Ag(111), I$_t$ = 2 pA, V$_b$ = 0.5 V, oscillation = 60 mV. (c-e) V$_{SP}$ maps of a diamond shaped 4-molecule island in three different charge states representing the N-1, equilibrium N, and N+1 states determined from fitting three segments to $\Delta$f(V) curves in b (5 nm x 5 nm). (h-j) shows the same sequence for a constant-height pixel-by-pixel grid on a ``Clover'' cluster, with set-point on NaCl(2ML)/Ag(111), I$_t$ = 2 pA, V$_b$ = 0.5 V, oscillation = 45 mV (4.5 nm x 4.5 nm).  Note that ``B'' sites correspond to sites 1 and 3, with ``A'' sites corresponding to sites 2 and 4. For the ``Clover'' cluster all 4 sites are equivalent. All scale bars 1 nm.}
    \label{fig:CPDmaps}
\end{figure}

Piece-wise parabolic curves were fit to the data to determine the position of the maximum of the parabola for each segment and the position of the jump. The position of the maximum gives the surface potential, and is representative of the local surface charge, with more positive values indicating more negative charge present. Maps of the surface potential obtained from each segment are shown in figure \ref{fig:CPDmaps} (c-e) for the ``diamond'' cluster and (h-j) for the ``clover'' cluster.  The equilibrium charge distribution corresponds to the curve fit segment around 0V bias, in d and i for ``diamond'' and ``clover'' respectively. For the ``diamond'' cluster, we see that the equilibrium charge distribution (Fig. \ref{fig:CPDmaps}d) is not uniform, with more negative charge accumulated on the two equivalent sites labelled ``B'' (1,3), and less on the two sites labelled ``A'' (2,4). This is consistent with an offset in the local chemical potential due to a difference in the polarization environment (estimated $\Delta E_P=0.023 eV$) as described in previous work \cite{Cochrane.201593h}. For the curve segments beyond the jumps, the surface potential becomes more uniform, indicating that the charge on each molecular site is similar. The symmetric ``clover'' cluster has a relatively uniform charge distribution in all three segments corresponding to electron removal (h), equilibrium (i) and electron addition (j) states.

Compared to the location of the SOMO/SUMO or Lower/Upper Hubbard states, the position of these jumps in the $\Delta f(V)$ appears closer to zero bias, particularly for the ``Diamond'' cluster where the jump corresponding to electron addition to the cluster occurs at $0.15\pm0.01$ V localized over the ``A'' sites, and electron removal at $-0.21\pm0.01$ V localized over the ``B'' sites. This difference of $0.36\pm0.01$ eV is considerably smaller the $\sim 1.4$ eV gap seen on the single molecule in STS corresponding to the Hubbard U. The localization of electron addition over the ``A'' sites, and electron removal over the ``B'' sites corresponds to the equilibrium charge distribution: ``B'' sites show more electron density where electrons are removed, while ``A'' sites show less electron density where electrons can be added. In contrast, the positions of the jumps seen in the ``Clover'' cluster are more consistent with the positions of states seen in the STS, with the electron addition occurring at $0.35\pm0.01$ V and electron removal at $-0.23\pm0.06$, still showing a reduction in the apparent U, with little variation between the 4 sites. 

The data on the ``Diamond'' cluster in particular raises some questions: why is the equilibrium charge distribution non-uniform, and what does this imply about the overall charge state of the cluster? What are the charge states when electrons are (transiently) added or removed? Also, while it makes sense from a Hubbard model perspective that the electrons avoid higher charge distribution on the ``B'' sites, and can be added to the ``A'' sites (and conversely electrons are pulled out of the regions with higher charge density), there is still clearly a repulsive potential to overcome for electron addition and removal. Further, while the molecules in this planar geometry are not expected to exhibit strong hybridization, there seems to be some propensity to distribute charge across the cluster. To answer these questions, and relate the jump positions back to physical model parameters, we built an extended Hubbard model that included different site energies (e.g. the differences in the ``A'' and ``B'' sites due to polarization of other molecules in response to a charge placed on those sites\cite{Cochrane.201593h}), and inter-site electrostatic interactions that depend on occupation.

\subsection{Model details}
The asymmetric ``Diamond" cluster of four PTCDA molecules over a silver substrate was modelled using the following Hamiltonian: 

\begin{equation}
\begin{split}
    \hat{H} &= -t \sum_{<i,j>\atop \sigma = \uparrow,\downarrow} (\hat{a}_{i,\sigma}^{\dagger}\hat{a}_{j,\sigma} + h.c.) + U\sum_{i=1}^4\hat{n}_{i,\uparrow}\hat{n}_{i,\downarrow}
    \\
    &\quad  - \sum_{\sigma = \uparrow,\downarrow}(t_{BB}\hat{a}_{3,\sigma}^{\dagger}\hat{a}_{1,\sigma} + t_{AA}\hat{a}_{4,\sigma}^{\dagger}\hat{a}_{2,\sigma} +h.c.) 
    \\
    &\quad - \epsilon_{B} (\hat{n}_1 + \hat{n}_3) - \epsilon_{A} (\hat{n}_2 + \hat{n}_4) 
    \\
    &\quad + V\sum_{<i,j>}\hat{n}_i\hat{n}_{j} + V_{BB} \hat{n}_{1} \hat{n}_3 + V_{AA} \hat{n}_2 \hat{n}_4
\end{split}
\end{equation}
The first two terms are the standard Hubbard model, describing nearest neighbor hopping of electrons between adjacent sites and on-site Coulomb repulsion between electrons on the same site. As we argue below, the detailed behavior of the cluster is actually determined by the additional terms included (not U), which describe diagonal hopping between the B sites 1,3 and between the A sites 2,4, respectively; the on-site energies $\epsilon_{A/B}= \epsilon + E_{P, A/B}$ at these sites, equal to the sum of the electron affinity $\epsilon$ of PTCDA  plus the corresponding polarization energy $E_{P,A}$ or $E_{P,B}$ from other nearby PTCDA molecules; and repulsions between electrons on neighbor sites and between the electrons on  A-A and B-B sites, respectively. These nearest-neighbour Coulomb terms are similar to those included to explore Nagaoka ferromagnetism \cite{Campbell.1988,Kollar.1996} on lattice models, but with the addition of distinct cross-cluster terms. As usual, $\hat{n}_{i\sigma} = a^\dag_{i\sigma} a_{i\sigma}$ is the number operator for electrons with spin $\sigma$ at site $i=1,\dots,4$.

Of the 9 parameters defining the Hamiltonian, we can set  $U = 1.4eV$, $\epsilon_{B} = 0.485 eV$ and $\epsilon_{A} = 0.462 eV$ based on prior single molecule  measurements and electrostatic calculations (see supplemental information). For the remaining 6 parameters, we expect that $t$ and $ t_{BB}> t_{AA}$ are in the few meV range, while $V>  V_{BB}> V_{AA}$ are in the $100meV$ range. These additional constraints come from the geometry of the diamond cluster. 

To further constrain these parameters, we need to know the minimum energy $E_N$ when there are a total of $N$ electrons in the cluster. This is found by considering all possible partitions $N=N_{\uparrow}+N_{\downarrow}$ of spin-up and spin-down electrons, where $0\le N_{\uparrow}\le N$. For each such partition, the size of the Hilbert space is $m = {4 \choose N_{\uparrow}}{4 \choose N_{\downarrow}}$.  The $m \times m$ matrix of the Hamiltonian above was constructed in the occupation number basis, and its eigenvectors and eigenenergies were calculated numerically. $E_N$ is then the minimum eigenenergy found amongst all these partitions.

We can then find the cluster ground-state energy $E_{N_g}$ as the global minimum between all these $E_N$ values for $N\le 4$.  $N_g$ is the number of electrons in the cluster in the ground-state. We note that the value of $N_g$ is not known from the experiment.

The minimum energy needed to add one electron to the cluster is 
\begin{equation*}
    E_{\Delta N = 1} = E_{N_g+1} - E_{N_g}
\end{equation*}
while the minimum energy to remove one electron is 
\begin{equation*}
    E_{\Delta N = - 1} = E_{N_g-1} - E_{N_g}
\end{equation*}
For the correct parameters, these must match the experimentally measured energy cost of adding ($0.15\pm0.01 eV$) and removing ($0.21\pm0.01 eV$) one electron determined from the jump positions in the ``Diamond'' cluster.



While in principle $E_{\Delta N = \pm 1}$ are functions of all remaining model parameters, in practice they only   depend significantly on $V, V_{AA}$ and $V_{BB}$ because $t, t_{AA}$ and $t_{BB}$ are smaller by an order of magnitude. A constrained optimization algorithm was used to identify the values of $V, V_{AA}$ and $V_{BB}$ that give the correct values of $E_{\Delta N =\pm 1}$,  under the constraint that $V \geq V_{BB} \geq V_{AA} > 0$, when for convenience we set $t = 20$meV, $t_{AA} = t_{BB} = 0$ eV (a careful exploration of the role of these hoppings is below). These values are found to lie on short curve segments inside the $V, V_{AA}, V_{BB}$ space, with different curves corresponding to different $N_g$ values.

Of all these possible values for $V, V_{AA}$ and $V_{BB}$, we then kept only those  that also produced the correct electron distributions. Specifically, we require that $\langle n_{B} \rangle > \langle n_{A} \rangle$, {\em i.e.} there is a higher average occupation on the B molecules (sites 1,3) than on the A molecules (sites 2,4) if $N = N_g$,  while  $\langle n_{A} \rangle > \langle n_{B} \rangle$ when $N = N_g + 1$, as these were key features of the experimental observations shown in Fig. \ref{fig:CPDmaps}. These two conditions are also rather insensitive to the value of $t$, because higher $t$ reduces the difference $\langle n_{A} \rangle - \langle n_{B} \rangle$ but does not reverse its sign.

\begin{figure}
    \centering
    \includegraphics[width=1\linewidth]{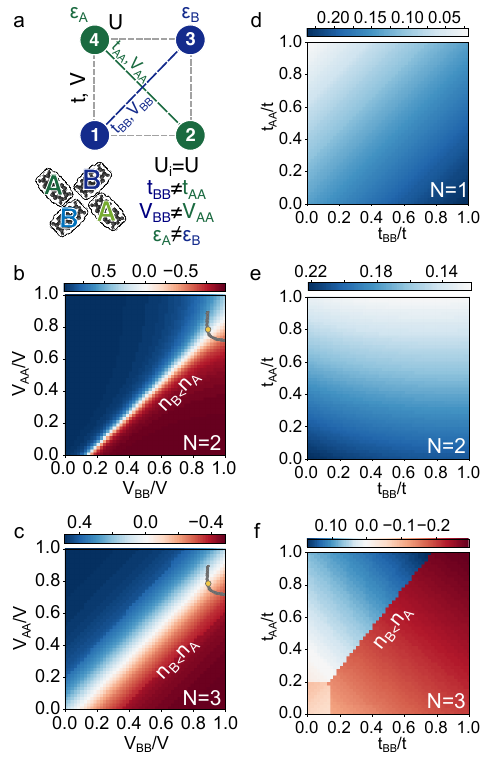}
    \caption{(a) 4-site extended Hubbard model mapping onto experimental system for ``diamond'' . The heat maps show the difference of the electron occupation $\langle n_{B}\rangle -\langle n_{A}\rangle$, as a function of $V_{AA}/V$ and $V_{BB}/V$ for fixed values of $V = 0.315 eV, t=0.02 eV$, and $t_{1,3}=t_{2,4}=0$ for $N=2$ (b) and $N=3$ (c),  as well as dependence on $t_{BB}/t$ and $t_{AA}/t$ for $N=1$ (d), $N=N_g=2$ (e, ground state), and $N=3$ (f), for fixed $U$, $V$, $V_{AA}$ and $V_{BB}$ determined from the data. In panels (b) and (c), the grey dots are the values giving the correct gaps $E_{\Delta N =\pm 1}$ while the orange circle is the final values given in the Discussion.}
    \label{fig:model}
\end{figure}

While in principle it is possible that $N_g = 2$ or $3$ (this value cannot be determined experimentally), these additional constraints eliminated all potential solutions with $N_g = 3$ because none of them had both acceptable electron density profiles and correct energy gaps. For $N_g = 2$, the solutions which have the correct energy gaps are shown as symbols in Fig. \ref{fig:model}(b) and (c).   The superimposed heat map shows the difference $\langle n_B\rangle -\langle n_A\rangle$, which must be positive (blue) in panel (b) and negative (red) in panel (c) for agreement with the experimental observations. Taken together, these conditions restrict the values of $V,~V_{AA},~ V_{BB}$ to a very narrow region, marked by the orange circle Fig. \ref{fig:model}(b) and (c).

With the ground state identified as having an overall occupation  $N_g=2$, and all other parameters fixed, we can now revisit the influence of the cross-cluster hopping parameters $t_{AA}$ and $t_{BB}$ on the site occupation for each cluster charge state (see Figure \ref{fig:model}, panels (d)-(f)). While the n.n. $t$ only serves to tune how localized the charges are on particular sites, the cross-cluster terms play a stronger role in determining the site occupation for each cluster charge state. For the $N_g=2$ ground state, increasing either $t_{AA}$ or $t_{BB}$ leads to more symmetric electron sharing across the cluster, while at low values, the charge is strongly localized on B sites 1 and 3 (see Fig. \ref{fig:model}e). For the $N=1$ electron removal state, the single electron is more strongly localized on B sites 1 and 3 as $t_{BB}$ is increased, while this effect is reduced as $t_{AA}$ is increased (see Fig. \ref{fig:model}d). Most interestingly, the $N=3$ electron addition state exhibits a quite dramatic switch in asymmetry where the charge on A sites 2 and 4 is greater than on the B sites 1 and 3 when $t_{BB}>t_{AA}$ (see Fig. \ref{fig:model}f). We expect this to be the case based on the short (long) distance between the B (A) sites, and indeed anticipate that the data for the diamond cluster is within this regime. This seemingly strange behaviour arises from a lowering of the cluster energy when an electron is delocalized over the B sites, while the other two electrons reside on the less connected A sites.

\subsection{Discussion}
\begin{figure*}
    \centering
    \includegraphics[width=1\linewidth]{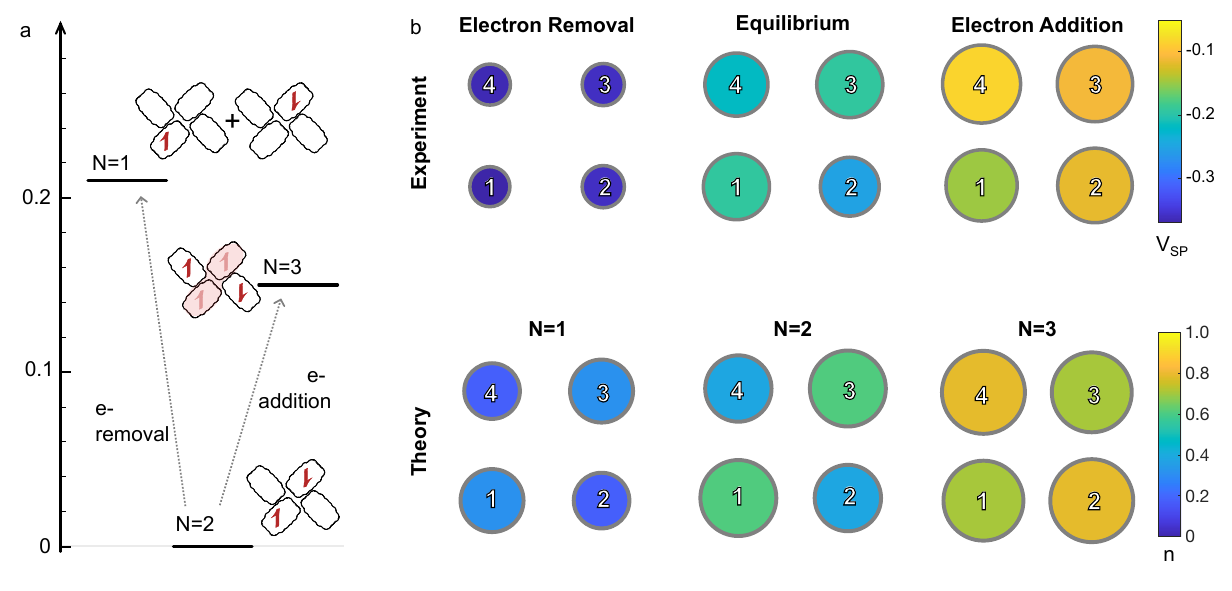}
    \caption{Charge occupation and excitations for the asymmetric ``diamond'' model and comparison to experimental data. (a) shows the charge excitations for electron addition and removal from the $N_g=2$ ground state. (b) shows the average surface potential reflective of the local charge for each site as the size of dot and intensity map, compared to the calculated fractional occupation as the size of dot and intensity map of each site for $N=1$, 2 and 3 from the model for $t=20meV$, $t_{AA}=t_{BB}=0$.}
    \label{fig:diamond-compare}
\end{figure*}

We can now closely examine the parameters extracted from the model in comparison with the data. The positions of the jumps in the $\Delta f(V)$ data constrained the model to a family of solutions with values of the potentials (with 1$\sigma$ standard deviations) $V=0.313\pm0.003 V$, $V_{BB}=0.287\pm0.016 V$, and $V_{AA}=0.256\pm0.0210 V$, for a value of $\epsilon=-0.4 eV$ relative to the $E_F$ of the Ag(111). These parameters are further constrained by requiring that $\langle n_{B}\rangle >\langle n_{A}\rangle$ for the $N=N_g=2$ solution while $\langle n_{B}\rangle < \langle n_{A}\rangle$ at $N=3$, giving $V=0.315\pm0.001 V$, $V_{BB}=0.281\pm0.002 V$, and $V_{AA}=0.247\pm0.006 V$. Both sets of values are close to those determined simply from the geometry of the clusters and an effective relative dielectric constant extracted from the reduced $U$ on the NaCl on Ag(111) substrate (see SI for details). We can now directly examine the partial occupation of each site for each charge state, which is dependent on all other parameters (except $U$, which merely avoids double occupancy). As the surface potential represents the local charge distribution, we can compare the experimentally measured $V_{SP}$ to the site occupancy from the model for reasonable parameters (see Fig. \ref{fig:diamond-compare}b) for all three charge states, taking the $N_g=2$ ground state to be the equilibrium observed near zero bias. We find that if we set $t=20$meV, and $t_{AA}=t_{BB}=0$meV, and keep all other parameters at the previously mentioned values, our model nicely reproduces the charge asymmetry at equilibrium, and subtler features such as the increased charge density on sites 2 and 4 in the $N=3$ case. Similar results can be obtained with small values of $t_{BB}>t_{AA}>0$, though these are indistinguishable experimentally without a quantitative conversion of surface potential to charge. Overall, these asymmetric charge distributions are driven by the site potentials (particularly in the $N_g=2$ case), cross-cluster repulsion, and the cross-cluster hopping parameters which lower energy with delocalization of a single electron only, as double occupancy is blocked.

In summary, a cluster of molecular anions was used to create a simplified case of a correlated electron system that is well described by an extended Hubbard model. In this large U regime, the jumps observed in the $\Delta f(V)$ spectroscopy correspond to transient changes in the overall charge state of the cluster and depend predominantly on the intersite potentials which are nearly an order of magnitude smaller than the Hubbard U, similar to the proportion of U to V for transition metals \cite{Hubbard.1963}. The (partial) electron occupation number of each site depends on the site potentials (which includes a uniform electron affinity, with a locally varying polarization energy), cross-cluster repulsion, and cross-cluster hopping parameters. These cross-cluster hopping parameters allow for delocalization of a single electron across two sites, lowering the overall energy of the system. As in examples of correlated systems where $t\sim U$, competition between the electrostatic potentials that drive localization and delocalization through wavefunction overlap lead to concerted motion of the electrons in the cluster with changes in the total electron occupation. Notably in this case, other than driving the system into a regime avoiding double occupation, the behaviour observed does not depend on $U$, which is often considered to sufficiently characterize the interactions in strongly correlated systems. Here instead it is the typically neglected inter-site potentials that drive the occupation and transition energies. 

The use of molecular systems to simulate Hubbard models offers several complementary aspects and potential advantages over existing platforms: (i) use of singly charged molecules provides a direct simulation platform for the fermionic Hubbard model; (ii) intermolecular spacing and connectivity/geometry can be tuned both by choice of substrate (e.g. different ionic salts with different lattice spacing or geometry), manipulation by the tip, or by including complementary synthons to drive self assembly and tune the local electrostatic environment; (iii) screening of both U and intermolecular potentials can be tuned by the layer thickness; and (iv) these systems are highly amenable to investigation by scanning probe techniques allowing not only read-out of the charge state of individual sites, but also the possibility to probe and control the spin-state using magnetically resolved techniques\cite{Czap.2019,Verlhac.2019,Chen.2023}, and the opportunity for local reporting via optical emission\cite{Kimura.2019,Dolezal.2022}.

\section{Methods}

Isolated PTCDA molecules on NaCl(2ML)/Ag(111) were obtained by thermally depositing NaCl (TraceSELECT $\geq$ 99.999\%, Fluka) on a clean Ag(111) single crystal substrate (Mateck GmbH) held at $\sim 80^{\circ}$C and subsequent deposition of PTCDA (98\%, Alfa Aesar) from a quartz crucible (Kentax GmbH) at $325^{\circ}$C onto a $\sim4.5$K surface. The substrate was then ``room temperature” annealed to form small islands by removing the sample plate from the cryostat and holding it in the wobblestick for $\sim5$ minutes\cite{Cochrane.201593h}.  This results in a variety of sizes and shapes of PTCDA clusters on NaCl driven by registry with the NaCl substrate and intermolecular interactions\cite{Burke.2009djs,Mativetsky.2007rgdl,Jia.2016,Cochrane.201593h,Scheuerer.2020}.

The measurements were performed in ultrahigh vacuum (UHV) at 4.3 K with a low temperature scanning probe microscope (LT-SPM, Scienta Omicron). A homemade qPlus sensor with a tungsten tip and separate gold wire contacting the tip for the tunneling current was used for all measurements, with f0 = 25500 Hz, Q ~ 40,000, and a sensitivity of $\sim 0.4$ pm/mV, determined by changing the oscillation amplitude and observing the current and the exponential dependence of tunneling current on tip-sample distance, known as the ``distance decay method". The bias was applied to the sample to further prevent coupling to the oscillation of the tuning fork. Constant height NC-AFM images and electrostatic force spectra were taken by positioning the tip over NaCl at 0.5 V and 2 pA using tunneling feedback, then turning off the feedback loop. Both carbon monoxide (CO) and PTCDA functionalized tips were used for high resolution imaging. EFS grid measurements were taken with a metallic tip with acquisition times typically around 100 milliseconds per voltage point and tip oscillations of 40 - 80 mV (16 - 32 pm); high resolution imaging was taken using tip oscillations of 20 - 60 mV (8 - 24 pm). All data acquired here was obtained with a Omicron Matrix controller and phase-locked loop.

\section{Acknowledgements}

The authors gratefully acknowledge funding from the Natural Sciences and Engineering Council (NSERC) of Canada, the Canadian Foundation for Innovation, and the Canada Research Chairs program (SB) for supporting this work. We would also like to acknowledge conversations with Alberto Nocera.

\bibliography{PTCDAmanuscript}

\end{document}